# Linear optical scheme for error-free entanglement distribution and a quantum repeater

Demetrios Kalamidas

*Institute for Ultrafast Spectroscopy and Lasers, City College of the City University of New York, 138$^{th}$ Street & Convent Avenue, New York, NY 10031, USA*

We present a linear optical scheme for error-free distribution of two-photon polarization entangled Bell states over noisy channels. The scheme can be applied to an elementary quantum repeater protocol with potentially significant improvements in efficiency and system complexity. The scheme is based on the use of polarization and time-bin encoding of photons and can perform single-pair, single-step purification with currently available technology.

The ability to share highly entangled multi-photon states between two or more distant parties is an essential requirement in most quantum communication protocols [1,2]. However, because of photon loss and decoherence [3], the delicate quantum states rapidly degrade within long-distance channels such as optical fibers. In theory, the main problems associated with the distribution of entanglement over noisy channels can be resolved with a quantum repeater protocol [4,5]. Yet, in practice, the quantum repeater presents serious technical challenges. In this Brief Report we describe a relatively simple and efficient linear optical scheme for error-free distribution of entanglement that can be applied to a quantum repeater protocol.

Consider scheme (a) of Fig.1. The source produces two-photon polarization entangled Bell states expressed as $|\Phi\rangle_{12} = \frac{1}{\sqrt{2}}(|H\rangle_1|H\rangle_2 + |V\rangle_1|V\rangle_2)$, where the kets represent the linear polarization state of a photon ('H' for horizontal and 'V' for vertical) and the subscripts denote each photon of the entangled pair (photon 1 going to Alice and photon 2 going to Bob). We assume that the two-photon states are created at definite times (i.e. from a pulsed source) with respect to a time reference. As shown, the source also includes two encoders (labeled 'E'), each of which is composed of an unbalanced polarization interferometer and a fast Pockels cell (PC). In the interferometers, $|H\rangle$ photons are transmitted by the polarizing beam splitters (PBSs) and propagate along the short path (S) while $|V\rangle$ photons are reflected by the PBSs and propagate along the long path (L). This causes the two amplitudes, $|H\rangle_1|H\rangle_2$ and $|V\rangle_1|V\rangle_2$, to be temporally separated. The two

alternative possibilities, "both photons are $|H\rangle$ and propagating in the *early* time-bin" or "both photons are $|V\rangle$ and propagating in the *late* time-bin", now comprise the superposition describing the two-photon state. The PCs implement bit-flips in the $|H\rangle/|V\rangle$ basis when they are activated. In the encoders, the PCs are activated only at the time when the 'late' ($|V\rangle$) components *are scheduled to arrive*, as coordinated by the time reference. When activated, the PCs effect the transformation $|V\rangle \to |H\rangle$. Hence, the two-photon state after the encoding is expressed as

$$|\Phi\rangle_{12} \xrightarrow{encoding} \tfrac{1}{\sqrt{2}}(|H\rangle_1^S |H\rangle_2^S + |H\rangle_1^L |H\rangle_2^L) \equiv |\Phi\rangle_{12}^E,$$

where the superscripts allude to the 'early' and 'late' time-bins that correspond to the S and L paths, respectively.

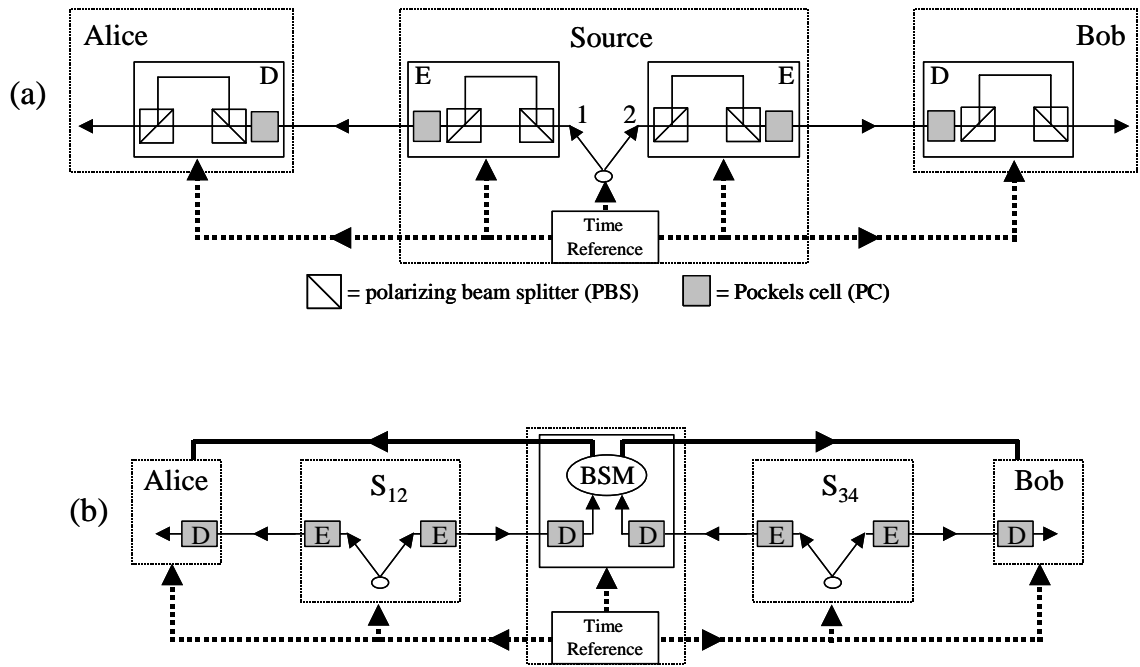

Fig.1. Error-free entanglement distribution and quantum repeater with linear optics. Scheme (a) depicts the main components required in distributing a maximally entangled photon pair over noisy channels. Scheme (b) outlines the method for incorporating scheme (a) into a quantum repeater protocol.

Suppose that the two noisy channels connecting Alice and Bob to the source are long-distance optical fibers. Apart from photon loss (which will be considered later) the principal factor inhibiting the distribution of the maximally entangled state is random birefringence arising from thermal fluctuations, vibrations, and imperfections of the fiber itself. This birefringence causes unknown transformations of the photon polarization state. For the encoding just described, where we assume the time-bins to be separated by several *nanoseconds*, the causes of birefringence are in a *steady-state*. This means that whatever unknown unitary operator, U, acts on the early component also acts on the late component of each photon. The operator U can be expressed by its action on the basis states:

$$|H\rangle \to e^{i\phi} \cos\theta |H\rangle + e^{i\chi} \sin\theta |V\rangle$$

$$|V\rangle \to -e^{-i\chi} \sin\theta |H\rangle + e^{-i\phi} \cos\theta |V\rangle,$$

which represents a general qubit transformation (excluding a global phase of no physical significance in this context). However, because of the encoding, the polarization state in both time-bins is $|H\rangle$ and the action of $U_i$ over long-distance channel 'i' (i=1,2) can be written as

$$|H\rangle_i^j \to e^{i\phi_i} \cos\theta_i |H\rangle_i^j + e^{i\chi_i} \sin\theta_i |V\rangle_i^j,$$

where the superscript denotes the time-bin (j=S,L). The evolution of the two-photon state from the source (after encoding) to just before Alice's and Bob's respective stations can be described by

$$|\Phi\rangle^{E}_{12} \xrightarrow[U_2]{U_1}$$

$$\tfrac{1}{\sqrt{2}}[(e^{i\phi_1}\cos\theta_1|H\rangle^{S}_1 + e^{i\chi_1}\sin\theta_1|V\rangle^{S}_1)(e^{i\phi_2}\cos\theta_2|H\rangle^{S}_2 + e^{i\chi_2}\sin\theta_2|V\rangle^{S}_2)$$

$$+(e^{i\phi_1}\cos\theta_1|H\rangle^{L}_1 + e^{i\chi_1}\sin\theta_1|V\rangle^{L}_1)(e^{i\phi_2}\cos\theta_2|H\rangle^{L}_2 + e^{i\chi_2}\sin\theta_2|V\rangle^{L}_2)]$$

$$\equiv |\Phi\rangle^{U_1,U_2}_{12}.$$

Alice and Bob possess decoders (labeled 'D') that are constructed from the same components used in the encoders. The decoders differ in that the PCs precede the polarization interferometers and, again coordinated by the time reference, are activated only at a time corresponding to the *scheduled arrival* of the 'early' components of the photons, effecting the transformation $|H\rangle^{S}_i \leftrightarrow |V\rangle^{S}_i$. The evolution of state $|\Phi\rangle^{U_1,U_2}_{12}$ through the decoders can now be described by

$$|\Phi\rangle_{12}^{U_1,U_2} \xrightarrow{decoding}$$

$$\tfrac{1}{\sqrt{2}}[(e^{i\phi_1}\cos\theta_1|V\rangle_1^{SL} + e^{i\chi_1}\sin\theta_1|H\rangle_1^{SS})(e^{i\phi_2}\cos\theta_2|V\rangle_2^{SL} + e^{i\chi_2}\sin\theta_2|H\rangle_2^{SS})$$

$$+(e^{i\phi_1}\cos\theta_1|H\rangle_1^{LS} + e^{i\chi_1}\sin\theta_1|V\rangle_1^{LL})(e^{i\phi_2}\cos\theta_2|H\rangle_2^{LS} + e^{i\chi_2}\sin\theta_2|V\rangle_2^{LL})]$$

$$= e^{i\phi_1}e^{i\phi_2}\cos\theta_1\cos\theta_2[\tfrac{1}{\sqrt{2}}(|H\rangle_1^{LS}|H\rangle_2^{LS} + |V\rangle_1^{SL}|V\rangle_2^{SL})] + |\Omega\rangle_{12}^{SS,LL},$$

where the superscripts indicate the paths traversed by the photons during encoding and decoding. There are now three time-bins for the arrival of a photon: 'too early', corresponding to SS; 'too late', corresponding to LL; 'intermediate', corresponding to SL or LS. From the last line of the above expression we readily observe that the maximally entangled state produced at the source is shared by Alice and Bob, despite its distribution over two noisy quantum channels, if both photons arrive within the intermediate time-bin. The state $|\Omega\rangle_{12}^{SS,LL}$ contains all the terms that involve one or both of the photons arriving too early or too late.

If Alice and Bob are equipped with time-gating devices, they can accomplish single-pair, single-step purification of the distributed entangled state by post-selection of those events wherein both photons are ultimately detected within the intermediate time-bin (all error-states having been rejected by way of this post-selection process). As is evident from the description of the protocol, the time reference coordinating distant parts of the system is a crucial element.

Concerning the performance of the protocol, the random variation of the parameters $\phi_i, \chi_i$ cannot induce errors and the uncorrupted maximally entangled state is obtained with probability $\cos^2 \theta_1 \cos^2 \theta_2$. This property is desirable since it means that for small values of $\theta_i$ the probability is close to 1. Allowing $\theta_i$ to vary randomly over their entire range (indicative of strong environmental influences) only results in a probability that tends (after many trials) towards the value $\frac{1}{2} \times \frac{1}{2} = \frac{1}{4}$, with *no* effect on the error-rejecting (purifying) capability of the scheme. This stands in stark contrast to typical multi-pair purification schemes that involve a threshold fidelity below which they fail, making them very susceptible to the strength of the environmental influences on the channel.

Scheme (a) can be applied, in a straightforward manner, to the construction of a linear optical quantum repeater, as shown in scheme (b) of Fig.1. In this scenario, two sources of polarization Bell states, $S_{12}$ and $S_{34}$, are used to connect Alice and Bob. The two-photon state from each source is subjected to the protocol of scheme (a) and, subsequently, one photon from each pair (photons '2' and '3') is used to execute a Bell-state measurement (BSM) wherein the two photons are processed only if they both arrive in the intermediate time-bin. The outcome of the BSM is then communicated to Alice and Bob via classical channels. Apart from the time-bin selection, this is just the procedure required for a standard quantum repeater protocol based on purification [6] and swapping [7], whose aim is to overcome the problems with entanglement distribution (due to exponential photon loss and decoherence) associated with increasing distance. Once more, the various procedures must be executed at definite times with respect to a reference. A precise and elaborate timing network will be required to synchronize the distant parts of the system. However, since timing is an integral part of any quantum repeater regardless of its embodiment, our set-up appears to offer significant advantages over other currently proposed designs. In this light, we consider the description and analysis of a linear optical quantum repeater given by Kok, Williams, and Dowling (KWD) [8].

KWD first explain the severe shortcomings of parametric down-converters as the source of two-photon entangled states, suggesting their replacement with near-deterministic 'double-photon guns' whose realization appears to be within reach [9,10]. They then go on to describe the other ingredients required in assembling a linear optical quantum repeater based on multi-pair purification and swapping: controlled-NOT (CNOT) operations, quantum nondemolition (QND) measurements, and Bell-state analyzers (all of which are probabilistic). In the KWD scheme, the probability of obtaining one maximally entangled (purified) photon pair between, say, Alice and the BSM station is given by

$$P_{pur}^{KWD} = (1-\gamma)\zeta p_S^5 \eta^8 p_{CNOT}^2 p_{QND},$$

where $\gamma$ (=0.5) quantifies the reduction in fidelity with increasing distance, $\zeta$ (=0.5) is the photon loss factor, $p_S$ (=0.9) is the probability of the source creating the two-photon state, $p_{CNOT}$ (=0.25) and $p_{QND}$ (=0.125) are the success probabilities for the CNOT operation and QND measurement, respectively, and $\eta$ is the detector efficiency. It is apparent that detector efficiency plays the dominant role in determining the effectiveness of the KWD scheme and, for $\eta = 0.3$ (0.8), $P_{pur}^{KWD} \approx 10^{-7}$ ($10^{-4}$). For the scheme in (b), where probabilistic CNOT operations and QND measurements are *not* required, the same probability is given by

$$P_{pur} = \zeta p_S \cos^2 \theta_1 \cos^2 \theta_2.$$

If we take $\cos^2\theta_1 = \cos^2\theta_2 = 0.5$ then $P_{pur} \approx 10^{-1}$, which is six (three) orders of magnitude greater than $P_{pur}^{KWD}$ with $\eta = 0.3(0.8)$. Furthermore, the requirement of only one double-photon gun per purification segment (instead of the five in the KWD scheme) enables the demands on this single source to be greatly relaxed. For instance, even if we reduce $p_S$ from 0.9 to 0.1, we still have $P_{pur} \approx 10^{-2}$. With regard to the BSM, both schemes are subject to the same limitations: success probability of 0.5 and two-fold coincidence detection. Therefore, the probability of successful entanglement swapping, after two pairs between neighboring segments have been purified, is $\eta^2/2$. Of course, these probabilities should be viewed only as 'reasonably good' theoretical estimates because although they include realistic values for channel loss, source efficiency, and detector efficiency, there will still be additional small amounts of loss and decoherence due to factors such as detector dark counts, mode-matching, imperfect unitary transformations, etc. that are associated with the various devices in place.

In conclusion, we have presented a linear optical scheme for error-free entanglement distribution and have described its role within an elementary (two segments, one swap) quantum repeater protocol. Our method relies on a novel use of polarization and time-bin encoding of photons to perform single-pair, single-step purification of polarization entangled states, evading the use of probabilistic CNOT operations and QND measurements. Compared with the KWD scheme, our method potentially offers dramatic improvement in both efficiency (up to six orders of magnitude increase in purification rate) and system complexity (i.e., one source per purification segment, instead of five sources and eight detectors). As with the KWD paper, our analysis does not address the issues of quantum memory and deterministic Bell-state sources, two necessary features of *any* quantum repeater. Of the two, long-lived quantum memory appears to present the greatest challenge, although exciting research is currently being conducted in this domain [11]. As noted by KWD [8], deterministic Bell-state sources may be available in the near term. Only when these two features become available can our scheme be considered for implementation within a quantum repeater protocol. In the meantime, the scheme may be worth testing in the context of error-free quantum teleportation and entanglement-based quantum key distribution in the range of up to tens of kilometers [3]. In such applications only one purification segment is required (no need of quantum memory) and conventional pulse-pumped parametric down-converters can be used to supply the two-photon Bell states (no need of near-deterministic double-photon guns).

Besides stressing the importance of a sophisticated timing network, the technical aspects involved in a real-world implementation of our scheme have not been touched upon in this paper, but they are essentially the same (time-bin entanglement, single-photon detection, coincidence detection, interferometer stability, etc.) as those that have been successfully realized in numerous linear optical quantum information protocols to date [1].

Partial financial support for this research was provided by NASA.